\documentclass[prx,twocolumn,showpacs,preprintnumbers,amsmath,amssymb,floatfix]{revtex4-1}

\usepackage{graphicx}
\usepackage{dcolumn}
\usepackage{bm}
\usepackage{xcolor}
\usepackage[utf8]{inputenc}
\usepackage{hyperref}
\usepackage{ulem}
{

\hypersetup{
    colorlinks=true,
    linkcolor=magenta,
    filecolor=magenta,      
    urlcolor= teal,
    citecolor=blue
}

\newcommand{\affA}{Department de F\'isica, Universitat Polit\'ecnica de Catalunya, Campus Nord B4-B5, E-08034, Barcelona, Spain}
\newcommand{\affAA}{Departament de F{\'i}sica Qu{\`a}ntica i Astrof{\'i}sica, Facultat de F{\'i}sica, Universitat de Barcelona, E-08028 Barcelona, Spain}
\newcommand{\affAAA}{Institut de Ci{\`e}ncies del Cosmos, Universitat de Barcelona, ICCUB, Mart{\'i} i Franqu{\`e}s 1, E-08028 Barcelona, Spain}
\newcommand{\affB}{Institut f\"ur Theoretische Physik, Leibniz Universit\"at Hannover, Appelstr. 2, 30167 Hannover, Germany}
\newcommand{\affE}{Zentrum f\"ur Optische Quantentechnologien, Fachbereich Physik, Luruper Chaussee 149, D-22761 Hamburg, Germany}
\newcommand{\affG}{Faculty of Physics, University of Warsaw, Pasteura 5, PL-02093 Warsaw, Poland}

\begin{document}

\title{Many-body bound states and induced interactions of charged impurities in a bosonic bath}




\author{Grigory E. Astrakharchik$^{1,2,3}$}
    \email{grigori.astrakharchik@upc.edu \\ luis.ardila@itp.uni-hannover.de}
 \author{Luis A. Pe\~na Ardila$^4$\textsuperscript{\textasteriskcentered}\thanks{luis.ardila@itp.uni-hannover.de}}
\author{Krzysztof Jachymski$^{5}$} 
\author{Antonio Negretti$^6$}
\affiliation{$^1$\affA}
\affiliation{$^2$\affAA}
\affiliation{$^3$\affAAA}
\affiliation{$^4$\affB}
\affiliation{$^5$\affG}
\affiliation{$^6$\affE}

\date{\today}


\begin{abstract}
Induced interactions and bound states of charge carriers immersed in a quantum medium are crucial for the investigation of quantum transport. Ultracold atom-ion systems can provide a convenient platform for studying this problem. Here, we investigate the static properties of one and two ionic impurities in a bosonic bath using quantum Monte Carlo methods. We identify three bipolaronic regimes depending on the strength of the atom-ion potential and the number of its two-body bound states: a perturbative regime resembling the situation of a pair of neutral impurities, a non-perturbative regime that loses the quasi-particle character of the former, and a many-body bound state regime that can arise only in the presence of a bound state in the two-body potential. We further reveal strong bath-induced interactions between the two ionic polarons.  Our findings show that numerical simulations are indispensable for describing highly correlated impurity models.
\end{abstract}

\maketitle

{\large \textbf{Introduction}}

Compound systems consisting of impurities immersed in a quantum medium are of fundamental importance in quantum many-body physics. A few relevant examples in the solid-state realm are the Kondo effect~\cite{anderson1970}, transport of heavy impurities in a Fermi liquid~\cite{Rosch1999}, and pair formation~\cite{Alexandrov2011,scalapino2018}. Dressing the impurity particle with the low-energy excitations of the medium can lead to the emergence of a quasi-particle called the polaron. Its physical realization in ultracold atomic setups offers a unique opportunity to dynamically control the system's parameters, such as the interaction strength~\cite{ArdilaPRA2015,ArdilaPRA2016,Luis2019,Levinsen2021}. Atom-ion quantum systems~\cite{HarterCP14,CoteAAMOP16,TomzaRMP} hold the promise to study polaron physics in the so-called strong-coupling limit~\cite{CasteelsJLTP11}, owing to the long-ranged character of the two-body impurity-bath interaction. Furthermore, ionic impurities are an excellent platform for studying transport phenomena, as the charge can be easily detected and dragged with an external electric field~\cite{dieterle2020transport}. In contrast to the neutral case, exotic transport properties due to macroscopic atomic dressing of the ion can be expected~\cite{GrossAP62}. Other quantum ion-atom-based simulation ideas include, e.g., the formation of lattice bipolarons with low effective mass~\cite{JachymskiPRR2020} and ion-induced interactions in Tomonaga-Luttinger liquids~\cite{Michelsen2019}. Furthermore, such setups can be relevant in the context of quantum simulation~\cite{Bissbort2013,Dehkharghani2017,JachymskiPRR2020}, quantum transport~\cite{CotePRL00,Mukherjee2019,Mostafa2019,ChristensenPRA2022}, as well as applications in quantum information processing~\cite{Hauke2010,Secker2016} and thermometry~\cite{oghittu2022}. A few experimental groups have recently attained the ultracold collisional regime in radio-frequency traps~\cite{Feldker2019,Schaetz_NAT21}. Alternatively, ions can be created in an ultracold gas by ionization of selected atoms from the bath~\cite{Kleinbach2018,Kroker2021}.

Since the first observation of a single ion in a radiofrequency trap~\cite{Toschek1980}, trapped ions have proven to be an excellent testbed to verify predictions of quantum theory as, e.g., quantum jumps~\cite{quantumjumpshh,quantumjumpsusa} and the Zeno effect~\cite{quantumzeno}, but also to trigger various fields of research and technology such as atomic clocks, quantum computation and simulation~\cite{rmp-ions,SchneiderRPP12,kokail2019self,daley2022practical}. Nowadays, tens of ions can be isolated and individually manipulated to implement quantum computing schemes and simulate spin models~\cite{monroe2021programmable}. Quantum circuits based on one- and two-qubit gates are routinely accomplished in laboratories~\cite{kielpinski2002architecture,egan2021fault,postler2022demonstration}. Ion logic gates usually require the ions to be sufficiently cold vibrationally. With the increasing complexity of the algorithms and thereby the number of required gates, this condition becomes harder to meet as the ions will inevitably be heated by the applied laser pulses. A possible solution to this issue is to use another quantum system as a coolant such that the ions are kept sufficiently cold to ensure fault tolerance. Atom-ion quantum mixtures are a prominent candidate here since ultracold gases easily reach sub-$\mu$K temperatures. Theoretical studies have shown that cooling of ions to the $s$-wave regime in the presence of micromotion can be made efficient by choosing a large ion-to-atom mass ratio~\cite{Cetina2012,KrychPRA15,Oghittu_PRA21}. Such studies, however, do not take into account the possibility of the formation of many-body bound states~\cite{CotePRL02,SchurerPRL17,Gregory2021,georg2021}, whose occurrence substantially affects the properties of the mixture. \\

In our previous study~\cite{Gregory2021}, we investigated the polaronic properties of a single ion immersed in a bosonic bath, identifying different regimes depending upon the presence of a two-body atom-ion bound state: a polaronic branch when it is absent, and a many-body bound-state (MBBS) branch when a two-body bound state is supported. The first regime is well described by a particle dressed by the low-energy excitations of the gas. Instead, the MBBS branch is characterized by clustering of atoms around the ion leading to a large effective mass, proportional to the number of bound bosons. The identified polaronic states cannot be described by the conventional Fr\"ohlich paradigm~\cite{Tempere2009,Grusdt2016,CasteelsJLTP11}, Bogolyubov theory~\cite{ShchadilovaPRL16,GrusdtPRA17,Ardila2021-1}, as well as field theoretical methods~\cite{georg2021,Ding2022}, since the system properties rely not only on the scattering length and the effective range of the two-body atom-ion interaction, but also on its long-range tail. Note that even for a neutral polaron beyond-Bogolyubov density modulation can be substantial~\cite{schmidt2022self}. Furthermore, the formation of a MBBS renders highly inhomogeneous the bath density in the vicinity of the ion. We have refined the previous models of ion-atom MBBS based on mean-field approach~\cite{GrossAP62,CotePRL02}. \\


In this work, we investigate the ground state properties of two ions in a bosonic bath utilizing quantum Monte Carlo techniques. A timely question is to explore mediated interactions between the impurities and understand to what extent analytical approaches effectively describe them. The interaction between quasi-particles is not only crucial for conventional and high-$T_{\mathrm{c}}$ superconductivity~\cite{SCALAPINO1995329,Alexandrov2011}, but it is also instrumental for developing quantum technologies with compound atom-ion systems such as quantum sensors~\cite{JachymskiPRL2018, WasakPRA2018,oghittu2022} and hybrid interfaces for information processing~\cite{Hauke2010,Secker2016}.\\

{\large \textbf{Results}}

Depending on the details of the two-body atom-ion interaction, we identify three following regimes, illustrated pictorially in Fig.~\ref{Fig:schematic-regimes}:

\begin{figure}
\begin{centering}
\includegraphics[width=\columnwidth]{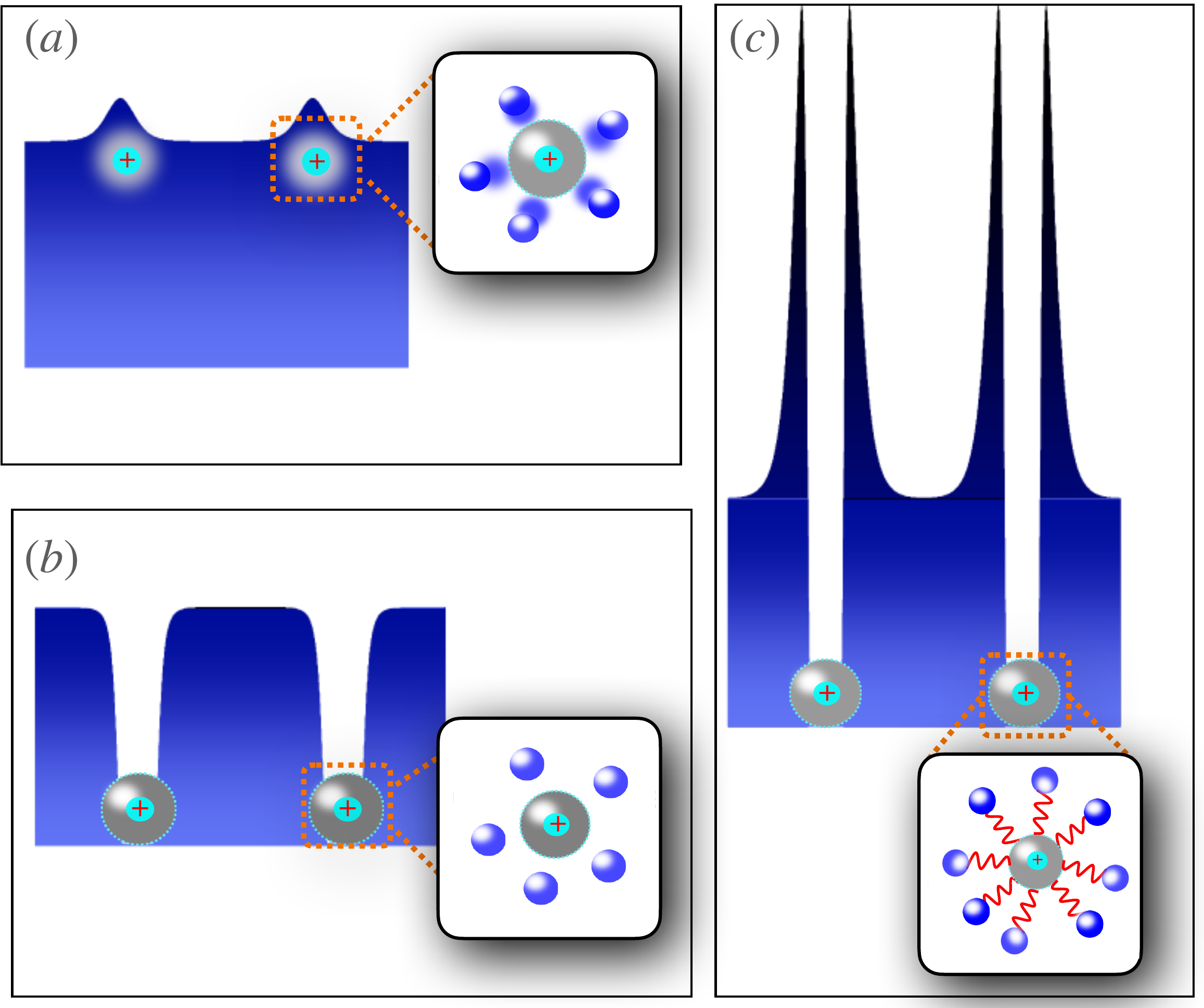}
\caption{\label{Fig:schematic-regimes}\textbf{Pictorial representation of the three identified regimes} of the ionic bipolaron consisting of two ions — cyan spheres labeled with “+” surrounded by a grey larger sphere representing the typical  range $R^{\star}$ — and a density of weakly interacting condensed bosons. The inset in each regime displays the atoms' distribution in the neighborhood of the ions. Panel (a) illustrates the weakly-interacting regime (i), where the local density modulation is small, thus permitting a quasiparticle description.  Panel (b) displays the regime (ii) with strong interaction having a hard core part. In this case, close to the ions the bath density vanishes. Panel (c) illustrates the regime (iii) in which the atom-ion interaction supports a two-body bound state (depicted by the wiggly red lines) participating in the formation of a many-body bound state. In this scenario, the bath density around the ion is strongly modified as many bosons are trapped in the vicinity of the ions.}
\end{centering}
\end{figure}

\begin{figure*}
\includegraphics[width=0.9\textwidth]{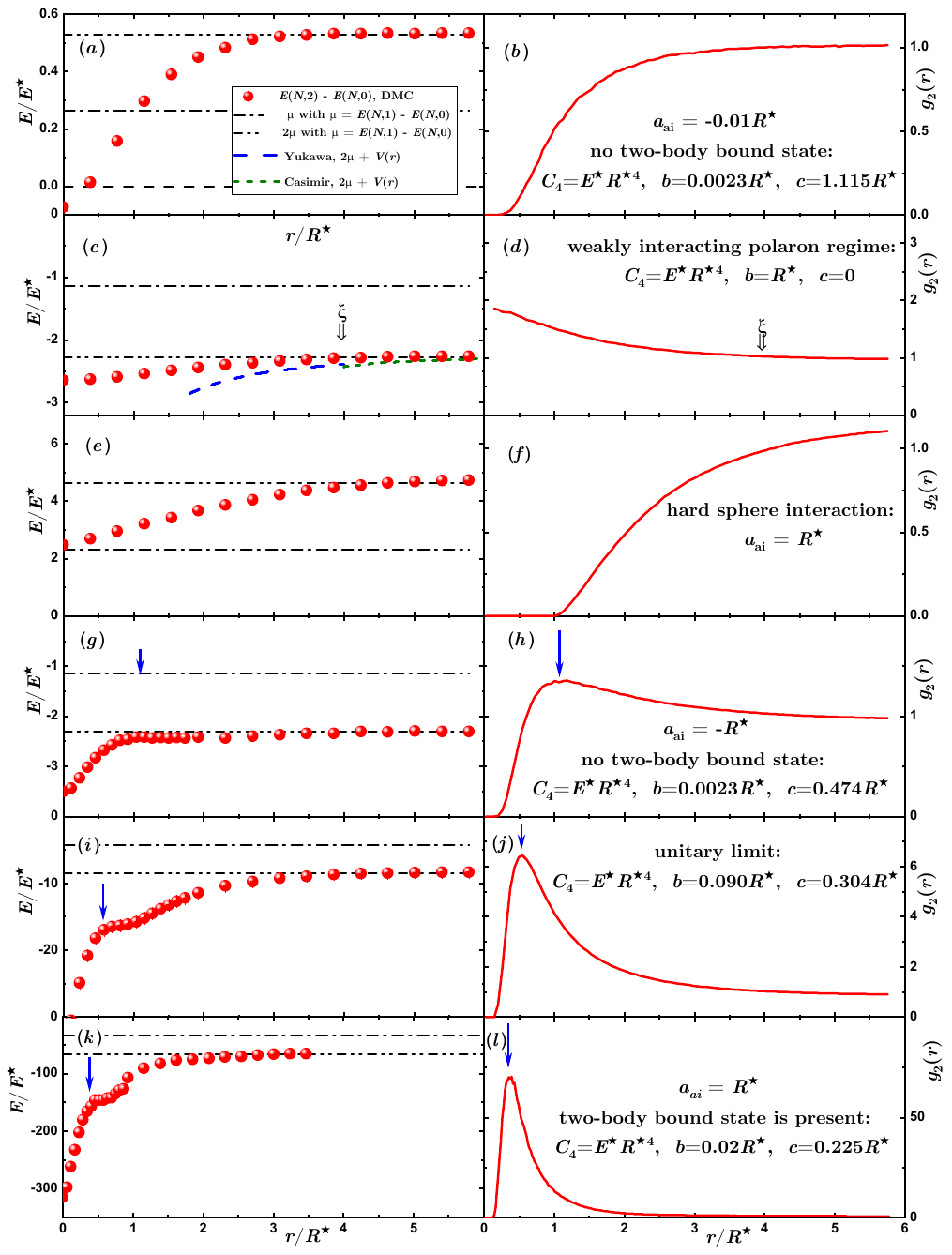}
\caption{\textbf{Polaron-polaron interactions and pair correlations functions.} Induced polaron-polaron interactions (left panels) and single polaron density profiles (right panels), which for a static ion are directly reflected by the ion-atom correlation function $g_2(r)$, as a function of their separation.
Panels show the results (a,b) in the weakly interacting regime;
(c,d) for $b=1$, $c=0$;
(e,f) for hard-sphere interactions;
(g,h) in the strongly-interacting regime;
(i,j) at unitarity;
(k,l) in the MBBS regime.
Symbols: results of QMC calculation,
black dash-dotted line: polaron chemical potential, 
black dash-dot-dotted lines: two polaron chemical potentials,
blue long-dashed line, Yukawa potential~(\ref{eq:Vind-shortR}) applicable at intermediate distances;
green short-dashed line, Casimir potential~(\ref{eq:Vind-largeR-Estar}) applicable at large distances. 
The blue arrows point to the location of the peak in the density.
}
\label{Fig:Eg2}
\end{figure*}

\begin{itemize}
\item[(i)] a perturbative (weak-coupling) regime;
\item[(ii)] a non-perturbative (strong-coupling) regime;
\item[(iii)] a many-body bound state regime. 
\end{itemize}
The weak-coupling regime, namely Fig.~\ref{Fig:schematic-regimes}(a), corresponds to the scenario in which the ion-induced density perturbation of the bath is small compared to the bath density at large distances from the ions, and therefore it can be treated perturbatively. In this case, we compare our many-body simulations for the induced interactions with the analytical results of Ref.~\cite{Ding2022}, and find a qualitatively similar behavior. However, for other parameters of the two-body potential we find large density modulations in the neighborhood of the ion(s), as depicted illustratively in Fig.~\ref{Fig:schematic-regimes}(b) and given quantitatively in right panels of Fig.~\ref{Fig:Eg2}. We refer to this situation as the non-perturbative regime (ii). The analytic theory assumes the validity of the Bogolyubov approximation for the condensate and neglects the contribution of the ion-atom bound states, taking into account only the exchange of phonons, such that it cannot be applied in this scenario. Finally, in regime (iii) the situation changes drastically because of the appearance of a two-body bound state in the atom-ion potential. Here, a so-called many-body bound state with hundreds of atoms is formed, a peculiarity of the compound atom-ion system. When two such ionic polarons are present, the nature of their interaction changes substantially. In Fig.~\ref{Fig:schematic-regimes}(c) we show the situation pictorially. Close to the ion, a low density region is created in the gas, while a cluster of bosonic atoms surrounds it with a peak in the bath density linked to the size of the molecular compound is formed. The scenario resembles the situation of ``snowballs" in helium~\cite{AtkinsPR1959}. In contrast to regimes (i) and (ii), the two-body atom-ion correlation function is highly non-monotonic with a peak value at some critical ion-ion separation, $R_{\mathrm{c}}$. A kink in the bath-induced interaction between the two polarons appears at this critical distance. Interestingly, below $R_{\mathrm{c}}$ the interaction increases enormously, that is, tens of times the energy scale of the atom-ion polarization potential (see, e.g., Fig.~\ref{Fig:Eg2}(k)). This behavior highlights the non-power-law character of the induced interaction at a short range.

\begin{figure}[h]
\begin{centering}
\includegraphics[width=\columnwidth]{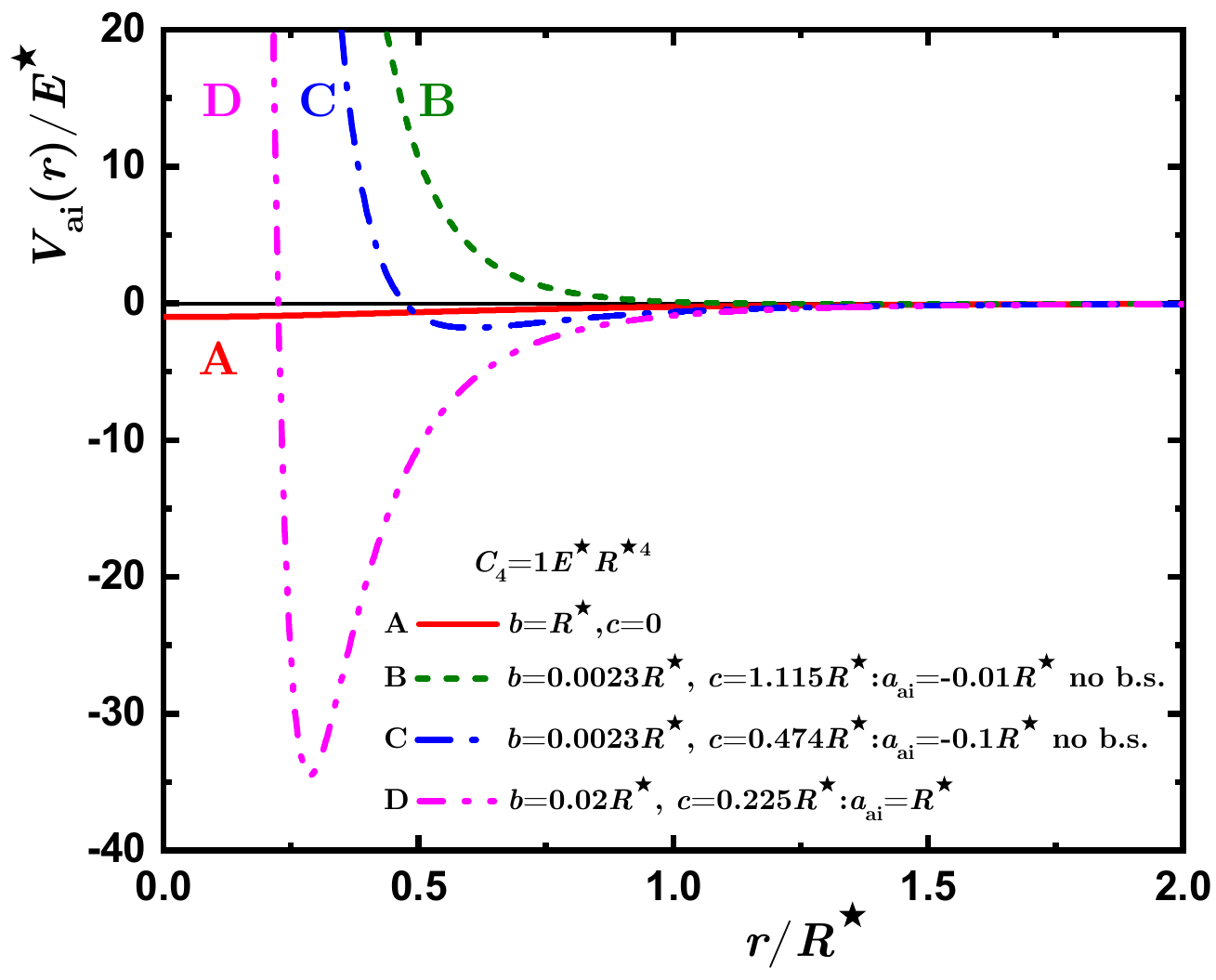}
\caption{\label{Fig:Vai}
\textbf{Atom-ion interaction potentials used in the many-body simulations.} 
Red solid line, $\mathrm{A}$  --- weakly-attractive model potential with no hard core;
green dashed line, $\mathrm{B}$ --- potential with a small negative scattering length leading to the absence of atom-ion bound states ($a_{\mathrm{ai}}=-0.01R^{\star}$);
blue dash-dot line, $\mathrm{C}$ --- potential with a large negative scattering length $(a_{\mathrm{ai}}=-0.1R^{\star})$ leading to the absence of an atom-ion bound state;
magenta dash-dot-dot line, $\mathrm{D}$ --- potential with a positive scattering length leading to the presence of an atom-ion bound state $a_{\mathrm{ai}}=R^{\star}$.
}
\end{centering}
\end{figure}

{\textbf{Model system} -- } \label{sec:System} The compound atom-ion system consisting of $N$ atoms and $N_{\mathrm{I}}$ pinned impurities is described by the following many-body Hamiltonian
\begin{align}
\label{eq:H}
\hat H\!=\sum_{n=1}^N \left[-\frac{\hbar^2\nabla^2_{\mathbf{r}_n}}{2 m} +\sum_{j=1}^{N_{\mathrm{I}}}V_{\mathrm{ai}}(\mathbf{r}_n\!-\!\mathbf{R}_j)\right]
\!+\!\sum_{n<j}^NV_{\mathrm{aa}}(\mathbf{r}_n\!-\!\mathbf{r}_j).\!
\end{align}
We assume that atoms obey Bose-Einstein statistics while statistics of impurities is not important if they are not allowed to move. (Hereafter, we denote the ion’s characteristics such as position and mass with capital Latin letters, while for atom ones we use small Latin letters. Furthermore, the bold symbol refers to three-dimensional vectors and cursive ones to their respective norms.) The first term represents the kinetic energy of the bosonic atoms of mass $m$, whereas $V_{\mathrm{aa}}(\mathbf{r}_n-\mathbf{r}_j)$ denotes the repulsive short-range atom-atom potential. 
The second term in Eq.~(\ref{eq:H}) describes the two-body atom-ion polarization potential, which possesses a long-range tail:
\begin{align}
\label{eq:Vai}
\lim_{r\to\infty}V_{\mathrm{ai}}(\mathbf{r}) \longrightarrow -\frac{C_4}{r^4}.
\end{align}
It is characterised by the length $R^\star = (2 m_{\mathrm{r}} C_4/\hbar^2)^{1/2}$ and energy scales $E^\star= \hbar^2 / [2m_{\mathrm{r}} (R^\star)^2]$, where $m_{{\mathrm{r}}} = m M/(m+M)$ is the reduced atom-ion mass. As an example, for the pair $^{23}$Na/$^{174}$Yb$^+$ we have $R^\star\simeq 129.85$~nm and $E^\star\simeq k_{\mathrm{B}}\times 0.71$~$\mu$K ($k_{\mathrm{B}}$ is the Boltzmann constant). For an atomic density $n = 6 \times 10^{13}$cm$^{-3}$, the mean inter-particle distance scales as $n^{-1/3}\simeq 2\,R^\star$, whereas the gas healing length $\xi=(8\pi na_{\mathrm{bb}})^{-1/2}\simeq 4R^{\star}$ with $a_{\mathrm{bb}}$ being the three-dimensional $s$-wave boson-boson scattering length, set by the gas parameter $na_{\mathrm{bb}}^{3}=10^{-6}$. 
As these lengths are all comparable, there is no separation of scales, and therefore short-range pseudopotentials cannot be used to replace the polarization potential, thus requiring the theory to take into account the atom-ion interaction potential explicitly. We model this interaction by the regularized potential~\cite{KrychPRA15}
\begin{align}
\label{eq:Vaireg}
V_{\mathrm{ai}}^{r}(\mathbf{r}) = -C_4\frac{r^2 - c^2}{r^2 + c^2} \frac{1}{(b^2 + r^2)^2}\, ,
\end{align}
with $b$, $c$ being regularization parameters that set the scattering length and the number of bound states in the system.
This choice of interaction potential has the benefit of retaining the long-range tail while also having a hardcore part and a simple form convenient for numerical and analytical calculations. As we are aiming for computing the ground state of the system, we choose to work in the range of parameters where the potential supports at most one bound state. This usually requires choosing rather large values of either the $b$ or $c$ parameter as compared to $R^\star$. In the following, we assume that the ions are static, i.e., they act as scattering centers for the bosons, and their separation is given by $\vert \mathbf{R}_1 - \mathbf{R}_2\vert = R$. Such a scenario is realized when heavy ions confined in a linear Paul trap are in interaction with light atoms. We can therefore omit the direct Coulomb interaction between them as well as their trapping potential. In a radiofrequency trap, the equilibrium separation between the closest ions along the crystal axis, say the $x$-axis, is approximately given by $R \simeq \alpha_{N_{\mathrm{I}}} \ell$ with $\ell^3 = e^2/(4\pi\epsilon_0 M\nu^2)$ and $\alpha_{N_{\mathrm{I}}}$ being a numerical factor that depends on the number of ions $N_{\mathrm{I}}$~\cite{James1998}. For large $N_{\mathrm{I}}$, it can be approximated as $\alpha_{N_{\mathrm{I}}} = 2.018 / N_{\mathrm{I}}^{0.559}$. Specifically, for two ions, we have $R \simeq 1.26\ell$. For $R = 1\,\mu$m and two $^{174}$Yb$^+$ ions a trap frequency of $\nu = 2\pi\cdot\,6$~MHz is required, while for 20~ions with approximately the same separation $\nu = 2\pi\cdot\,1$~MHz is needed. Note that related studies discussed the band structure of a single atom in a potential landscape generated by a chain of static ions~\cite{NegrettiPRB14,Marta2018,Marta2020}.\\ 

It is anticipated that the pinning of the impurities enhances the sharpness of density perturbations as compared to the situation in which the impurities are mobile. For example, this is known from the problem of an impenetrable one-dimensional gas in the presence of an impurity which is either mobile or pinned. The impurity profile shows Friedel oscillations in both cases, although the amplitude of the oscillations is larger in the pinned case~\cite{GirardeauWright2000}. Qualitatively, this can be understood in terms of a stronger interference pattern when an incident particle is bounced back from a non-moving impurity as compared to the situation in which the impurity can move. \\

{\textbf{Analytical expressions} --}
\label{sec:formulae}In Ref.~\cite{Ding2022}, the regularized potential~(\ref{eq:Vaireg}) has been utilized to predict induced ion-ion interactions. There, the impurity-bath interaction in second quantization is described by
\begin{align}
V_{\mathrm{ib}}(\mathbf{R}) = \sum_{\mathbf{k,q}} V_q \hat c_{\mathbf{k} + \mathbf{q}}^\dag \hat c_{\mathbf{k}} \left(1 + e^{- i \mathbf{q} \cdot \mathbf{R}}\right),
\end{align}
where $\mathbf{R}$ denotes the separation among the two ions, $V_q$ is the Fourier transform of the atom-ion potential, $\hat c_{\mathbf{k}}^\dag$ ($\hat c_{\mathbf{k}}$) denotes the creation (annihilation) operator of a boson with momentum $\mathbf{k}$. The theory implicitly assumes that the boson-boson interaction is sufficiently weak to make the Bogolyubov theory applicable, i.e., $n\, a_{\mathrm{bb}}^3\ll 1$ everywhere with $n$ being the local gas density. The induced interaction between ions $V_{\mathrm{ind}}(R)$ is obtained by calculating the energy shift due to their presence in the bath.

\label{sec:weak}

For large distances, i.e., $R\gg b, c, \xi$, it has been shown that 
\begin{align}
\label{eq:Vind-largeR-Estar}
V_{\mathrm{ind}}(R)=-\frac{\pi}{4}\frac{1}{a_{\mathrm{bb}}b}\frac{b^{2}+2bc-c^{2}}{(b+c)^{2}}\left(\frac{1}{R}\right)^{4}\, ,
\end{align}
where the length and energy units of $R^\star$, $E^\star$ have been used. The interaction~(\ref{eq:Vind-largeR-Estar}) has the same dependence on the distance as the atom-ion polarization potential~(\ref{eq:Vai}), but with varying sign which can be tuned by choosing proper combinations of the parameters $b$ and $c$. Due to its long-range power-law decay, we will refer to it as Casimir interaction. Instead, in the short distance limit, namely when $b, c \ll R\le \xi$, the bath-induced interaction becomes (again using $R^\star$, $E^\star$ units)
\begin{equation}
\label{eq:Vind-shortR}
V_{\mathrm{ind}}(R)=-\frac{\pi^{3}}{2}n\frac{1}{b^{2}}\frac{(b^{2}+c^{2})^{2}}{(b^{2}-c^{2})^{2}}\exp\left(-4\sqrt{\pi na_{\mathrm{bb}}}R\right),
\end{equation}
which has the form of a Yukawa interaction that is also obtained for neutral impurities in a condensate~\cite{Bruunprx2018,Luis2018}.\\ 


{\textbf{Weak coupling regime} --} The weak coupling regime is commonly associated with small values of the scattering length as compared to the inter-particle distance. This typically corresponds to a situation in which the energy shifts are small and the impurities only slightly distort the shape of the host gas. An important feature of our treatment of the ion impurity is that even for small values of the scattering length, there is an impenetrable wall in the atom-ion potential located at a relatively large distance $\sim R^\star$, as depicted in Fig.~\ref{Fig:Vai}. Thus, even though the energy shift might be small, the bath density remains strongly perturbed. This feature has strong consequences for the induced interaction between two polarons. \\

Figure~\ref{Fig:Eg2}(a) shows the QMC prediction for the polaron-polaron induced interaction obtained for a small value of the atom-ion scattering length $a_{\mathrm{ai}}=-0.01R^\star$, modelled by the potential  with parameters equal to $C_4=E^\star R^{\star4}$, $b=0.0023R^\star$, $c=1.115R^\star$ -- see also curve (B) in Fig.~\ref{Fig:Vai}. Note that our choice of parameters in this case is rather unphysical, as it leads to a hard wall at distances $R\sim R^\star$, but it highlights the role of the details of the regularization. In this case the energy scale of the induced interaction is set by the energy shift of a single impurity (i.e. polaron chemical potential $\mu$) and the spatial scale is set by the atom-ion potential range $R^\star$. While the induced interaction $V_{\mathrm{ind}}(r)$ itself is rather weak, perturbative expressions~(\ref{eq:Vind-largeR-Estar}-\ref{eq:Vind-shortR}) cannot be applied to this case. Mathematically, the amplitude of the Casimir potential for small values of $b$ diverges as $V_{\mathrm{ind}}(r)/E^\star\propto R^\star/b \approx 4.3\times 10^2$ and for the Yukawa potential as $V_{\mathrm{ind}}(r)/E^\star\propto (R^\star/b)^2 \approx 1.9\times 10^5$. As a result, the predictions of perturbative expressions would fall out of scale. The physical reason for the failure of the perturbative theory is that the density profile [see Fig.~\ref{Fig:Eg2}(b)] is completely voided at short distances, where the interaction potential is described by a hard wall (see the blowup of $V_{\mathrm{ai}}(r)$ in Fig.~\ref{Fig:Vai}). This violates the perturbative assumption used to derive the induced interactions between polarons.\\

In order to test the correctness of the analytic expressions for induced polaron-polaron interactions in the perturbative regime such that assumptions of both methods match each other, we perform calculations for the following parameter of the model atom-ion potential~(\ref{eq:Vaireg}), $C_4=E^\star R^{\star4}$, $b=R^\star$, $c=0$. While such a choice of parameters leads to no hard-core part of the interaction, the resulting potential (depicted by the curve (A) in Fig.~\ref{Fig:Vai}) is not perturbing the bath of atoms too strongly, and therefore the weakly-interacting polaron regime is realized. Figure~\ref{Fig:Eg2}(c,d) shows the induced polaron-polaron interactions in the perturbative regime and the polaron density profile. A reasonable agreement with the analytical predictions is found for the induced interactions. Namely,  for large distances, $r\gtrsim \xi$, the decay is compatible with a slow power-law characteristic of the Casimir effect.
Instead, at shorter distances, $R^\star\ll r \lesssim \xi$, there is a qualitative agreement in the shape similar to the fast-decaying Yukawa one typical to Bogolyubov theory. 
Still, there remains a quantitative difference with the perturbative expressions.

\begin{figure*}
\begin{centering}
\includegraphics[width=0.6\textwidth]{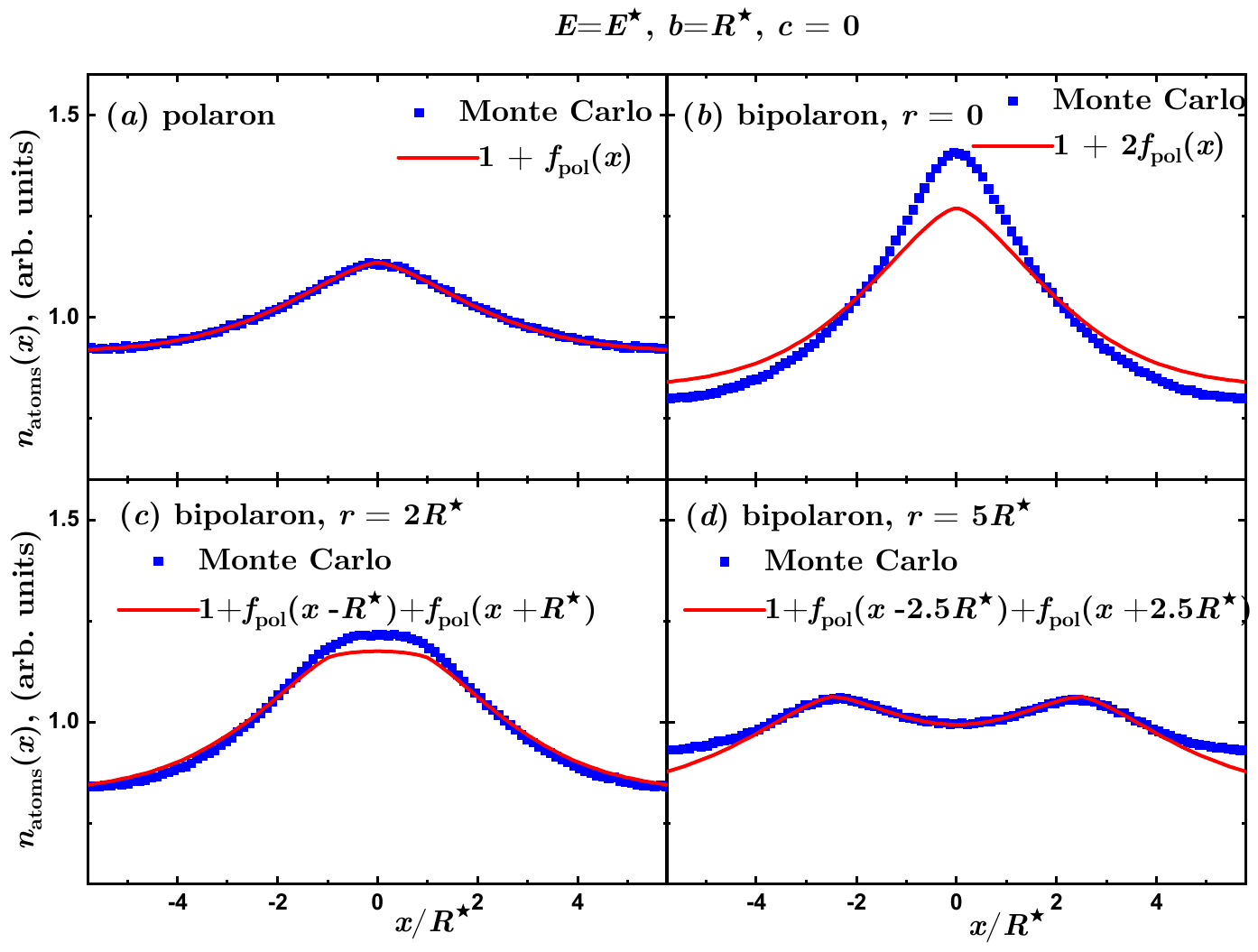}
\caption{\label{Fig:profile:c=0}
\textbf{Bipolaron density profiles at weak coupling.} Density profiles projected onto a line denoted by the $x$-axis connecting two impurities in the regime of weak interactions (arbitrary units).
Symbols: Blue squares denote the results of QMC calculations, while the solid line is a fit to the defect in the projected atom density
$n_{\mathrm{atoms}}(x)=1+f_{\mathrm{pol}}(x)$ 
with $f_{\mathrm{pol}}(x) = \Delta n
+A\exp[-(r/\sigma)^p]$, where $\Delta n,A,\sigma,p$ are fitting parameters.
Panel (a): Single ion (polaron) density profile used to obtain the fit $f_{\mathrm{pol}}(x)$.
Panels (b,c,d): Two ions (bipolaron) density profile for distances between them equal to $0,\,2R^\star,\,5R^\star$ as compared to the prediction for two non-interacting polarons located at the same ion positions.
}
\end{centering}
\end{figure*}

It is instructive to study how the bipolaron density profile depends on the distance between impurities. In the bipolaron case, the density profile no longer has a spherical symmetry of the single polaron [Fig.~\ref{Fig:Eg2}(d)]. For convenience, we project the atom density onto a single line connecting the two impurities in the bipolaron case and an arbitrary line passing through the impurity in the single polaron case. This results in the density profiles $n_{\mathrm{atom}}(x)$, which depend on one coordinate (denoted by $x$) as shown in Fig.~\ref{Fig:profile:c=0} for characteristic distances between the two ion impurities. The actual amplitude of $n_{\mathrm{atom}}(x)$ depends on the integration volume and hence arbitrary units are used on the vertical axis.\\

In the case of a single impurity, shown in Fig.~\ref{Fig:profile:c=0}(a), there is a mild density increase around the vicinity of the polaron. Its shape can be well fitted with a Gaussian-like profile. Panels Fig.~\ref{Fig:profile:c=0}(b,c,d) report the bipolaron density profile for three different separations between the impurities: $r=0, \,2R^\star,\,5R^\star$, and show the density profile of two non-interacting polarons separated by $r$. The bipolaron density profile recovers the density of two non-interacting polarons placed at large separation. This conclusion agrees with Bogolyubov's theory in which the induced interaction is small for distances large compared to the healing length, $r\gg\xi$. On the other hand, an enhancement in the density is visible for smaller and comparable distances, elucidating the attractive character of the induced interactions between polarons. For $r=0$ the density profile is a single pinned impurity having twice a stronger interaction strength. \\


{\textbf{Hard-sphere impurity} --}\label{sec:HS} In order to investigate further the role of the excluded region in the atom-ion interaction potential, i.e., in the region of the barrier, we perform simulations by considering a hard-sphere potential defined as $V_{\mathrm{ai}} = +\infty$ for $|{\bf r}|<a_{\mathrm{ai}}$ and zero otherwise. The atom-ion $s$-wave scattering length $a_{\mathrm{ai}}$ is then given by the size of the hard sphere. The QMC results are shown in Fig.~\ref{Fig:Eg2}(e,f). The polaron density depicted in Fig.~\ref{Fig:Eg2}(e) is completely depleted for distances $r<R^\star$. This fact has several important consequences: first, for zero separation between two hard spheres, the excluded volume remains exactly the same as for a single impurity and two overlapping hard spheres, and the system energy matches with $E(N,2) = E(N,1)$. This allows us to find the value of the induced interactions within the Born-Oppenheimer approximation for zero separation as $V_{\mathrm{ind}}(r=0) = E(N,2) - E(N,0) - 2\mu = -\mu$. Second, the amplitude of the induced polaron-polaron interaction is known exactly and it is given by the polaron chemical potential in that case. Within perturbation theory, the bipolaron shift energy would be equal to two polaron shift energies, while for hard spheres, both shifts are equal. The bipolaron energy $E(N,2)-E(N,0)$ is a continuous function, which goes from $\mu$ at $r=0$ to $2\mu$ at $r\gg\xi$, as shown in Fig.~\ref{Fig:Eg2}(e). \\

\begin{figure*}
\begin{centering}
\includegraphics[width=0.6\textwidth]{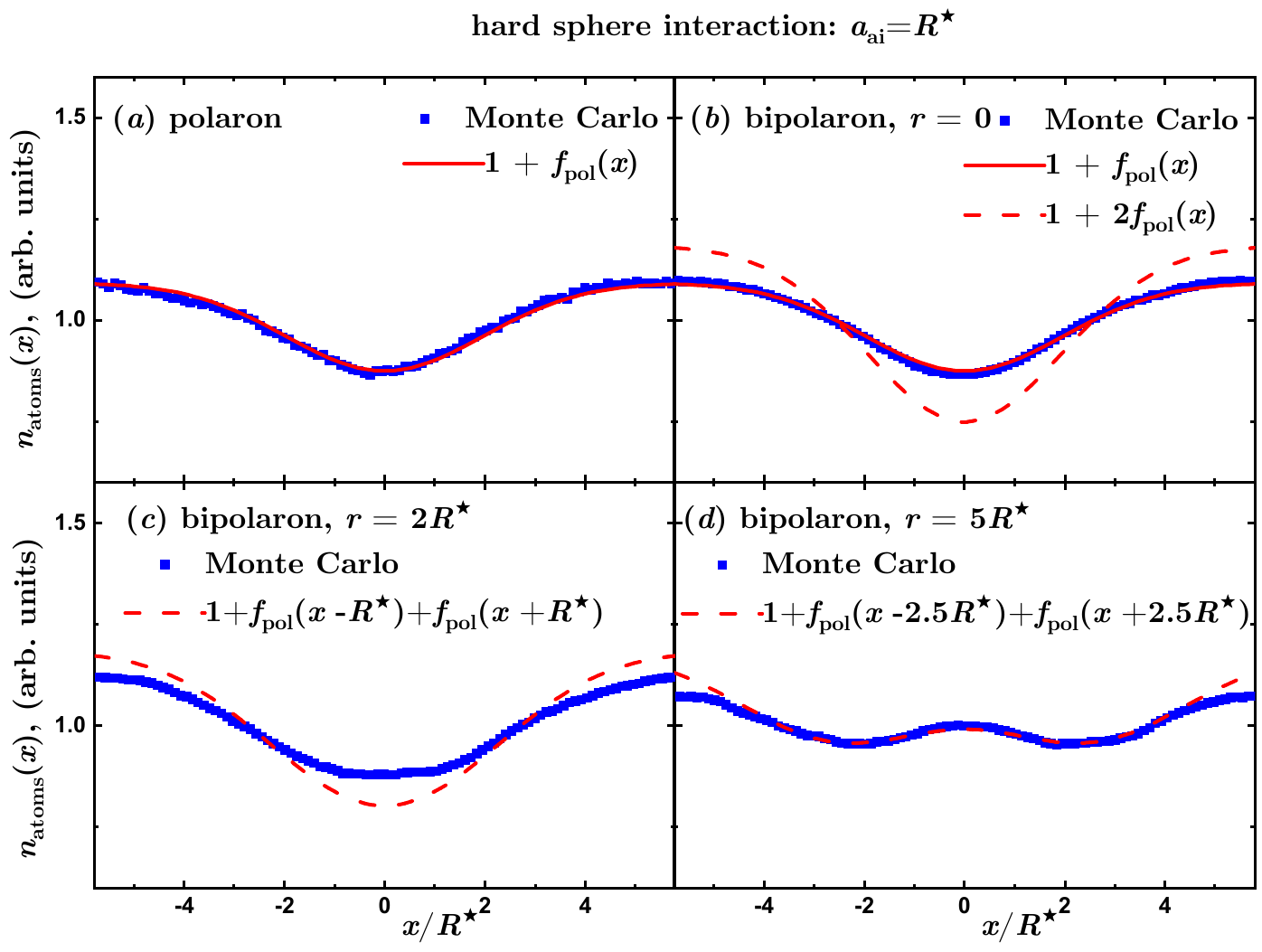}
\caption{\label{Fig:profile:HS}
\textbf{Bipolaron density profiles for a hard sphere potential.} Density profiles projected onto a line denoted by $x$-axis connecting two impurities for hard spheres (arbitrary units).
Symbols: Blue squares denote the results of QMC calculations, while the
solid line is a fit to the defect in the projected atom density $n_{\mathrm{atoms}}(x)=1+f_{\mathrm{pol}}(x)$ 
with $f_{\mathrm{pol}}(x) = \Delta n
+A\exp[-(r/\sigma)^2]$, where $\Delta n,\,A$, and $\sigma$ are fitting parameters. 
The dashed lines represent the estimation of the density of non-interacting polarons, 
$1+f_{\mathrm{pol}}(x+r/2)+f_{\mathrm{pol}}(x-r/2)$.
Panel (a): Single ion (polaron) density profile used to obtain the fit $f_{\mathrm{pol}}(x)$. Panels (b,c,d): Two-ion (bipolaron) density profile for the distance between ions equal to $0,\,2R^\star,\,5R^\star$ as compared to the prediction for two non-interacting polarons located at the ion positions.
}
\end{centering}
\end{figure*}

Typical bipolaron density profiles are presented in Fig.~\ref{Fig:profile:HS}. The repulsive hard-sphere potential leads to a depletion of the density around the impurity. Let us note that for the hard-core potential, it holds $2V_{\mathrm{HS}}(r)=V_{\mathrm{HS}}(r)$, that is, the potential experienced by the bath is the same for both a single impurity and two impurities separated by zero distance. As a consequence, the polaron density profile [Fig.~\ref{Fig:profile:HS}(a)] is exactly the same as the bipolaron density profile for $r=0$ [Fig.~\ref{Fig:profile:HS}(b)]. To a certain extent, this is the least perturbative case since the atom density is totally voided around the impurities. Notwithstanding, for a large separation $r$ between the impurities, the atom density is roughly equal to the densities of two independent polarons, signaling that the induced interactions are weak for such values of $r$.\\

{\textbf{Strong-coupling regime} --}
\label{sec:strong} A characteristic feature of ionic impurities is the possibility of realizing the strongly-interacting regime. Here we consider a large value of the atom-ion scattering length as well as a situation in which the atom-ion interaction does not support a bound state so that the scattering length is negative. A characteristic example of the (induced) polaron-polaron interaction potential in this regime is displayed in Fig.~\ref{Fig:Eg2}(g). A qualitative difference with respect to the previous two regimes (i.e., weak interaction and hard-spheres) is that the polaron density profile is no longer monotonous as it acquires a peak at a distance set by the potential range $R^\star$, see Fig.~\ref{Fig:Eg2}(h). 
The voiding of the single polaron density profile at short distances is due to the hard wall short-range repulsive part present in the potential and this feature is shared with the density profile obtained for hard sphere impurity, Fig.~\ref{Fig:profile:HS}. 
Instead, the peak is formed only for atom-ion potential and only for strong interactions, caused by the long-range attractive part of the potential.

Unlike the neutral polarons, the induced interaction potential also displays a non-monotonous behavior. Notably, the position of the peak in $V_{\mathrm{ind}}(r)$ coincides with the position of the peak in the density profile. We elaborate on this effect in the following paragraphs, where it is much more evident. Note that the strong-coupling regime differs from the previous two scenarios by a larger energy shift. \\


{\textbf{Unitary limit} --}
\label{sec:UNI} The most strongly interacting regime associated with $s$-wave scattering is the unitary limit in which the atom-ion scattering length diverges, $a_{\mathrm{ai}}\to \infty$. Analytically, such a regime is challenging due to the absence of a small parameter. Thus, it is instructive to study the ion bipolaron at unitarity. Figures~\ref{Fig:Eg2}(i,j) show the results obtained for unitary interactions. Already at the level of the single polaron case, atoms create a many-body bound state around the impurity as signaled by the presence of a very high peak in the density profile, see Fig.~\ref{Fig:Eg2}(j). The characteristic length at which the maximum appears in the polaron density profile is set by $R^\star$, and the induced ion-ion interactions have a spatial feature at that point, see Fig.~\ref{Fig:Eg2}(i). The bipolaron energy becomes an order of magnitude larger manifesting the formation of a deeply bound state which can be already understood at the single atom level. Moreover, for large distances between the impurities resonantly interacting with the host bath, a bound state has vanishing energy. For short separations, the potential landscape drawn by the two ions has an amplitude that is twice larger, leading to a formation of an atom-ion-ion bound state. Adding other atoms populates this bound state further, lowering the energy. 


{\textbf{Many-body bound state regime} --}
\label{sec:MBBS} The situation in which an ion impurity differs the most from neutral polarons is characterized by the formation of many-body bound states. Such a regime is reached when the two-body atom-ion problem supports a bound state. The properties of the system are dominated by the presence of a many-body bound state which acquires a large population (hundreds of atoms). On the single impurity level, this is manifested by the correlation function reaching higher values than the equilibrium gas density [see Fig.~\ref{Fig:Eg2}(l)]. The energy shift for the two ions becomes significantly larger than in all previously considered regimes, exceeding the bare sum of two single polaron shifts multiple times as depicted in Fig.~\ref{Fig:Eg2}(k). Moreover, in this regime, the cloud distortion impacts the shape of the induced interaction as well. At the point of maximal density, marked by the blue arrow in Fig.~\ref{Fig:Eg2}(k,l), the energy shift features a kink and stops growing rapidly with increasing separation. We interpret this effect as follows: at short distances, the two ions attract many bosons, leading to sizeable binding energy. The impurities effectively cooperate between them. Namely, the distance between the charges is increased and the local atomic density around them grows, still trapping atoms more strongly than a single ion would be able to. This is responsible for bending the curve in the vicinity of the blue arrow, which shows the maximum of the correlation function. The effect gets weaker at larger separations as the correlation function for a single impurity drops back to lower values, and thereby the mutual response between the impurities in attracting the bosons is inhibited. \\

\begin{figure*}
\begin{centering}
\includegraphics[width=0.6\textwidth]{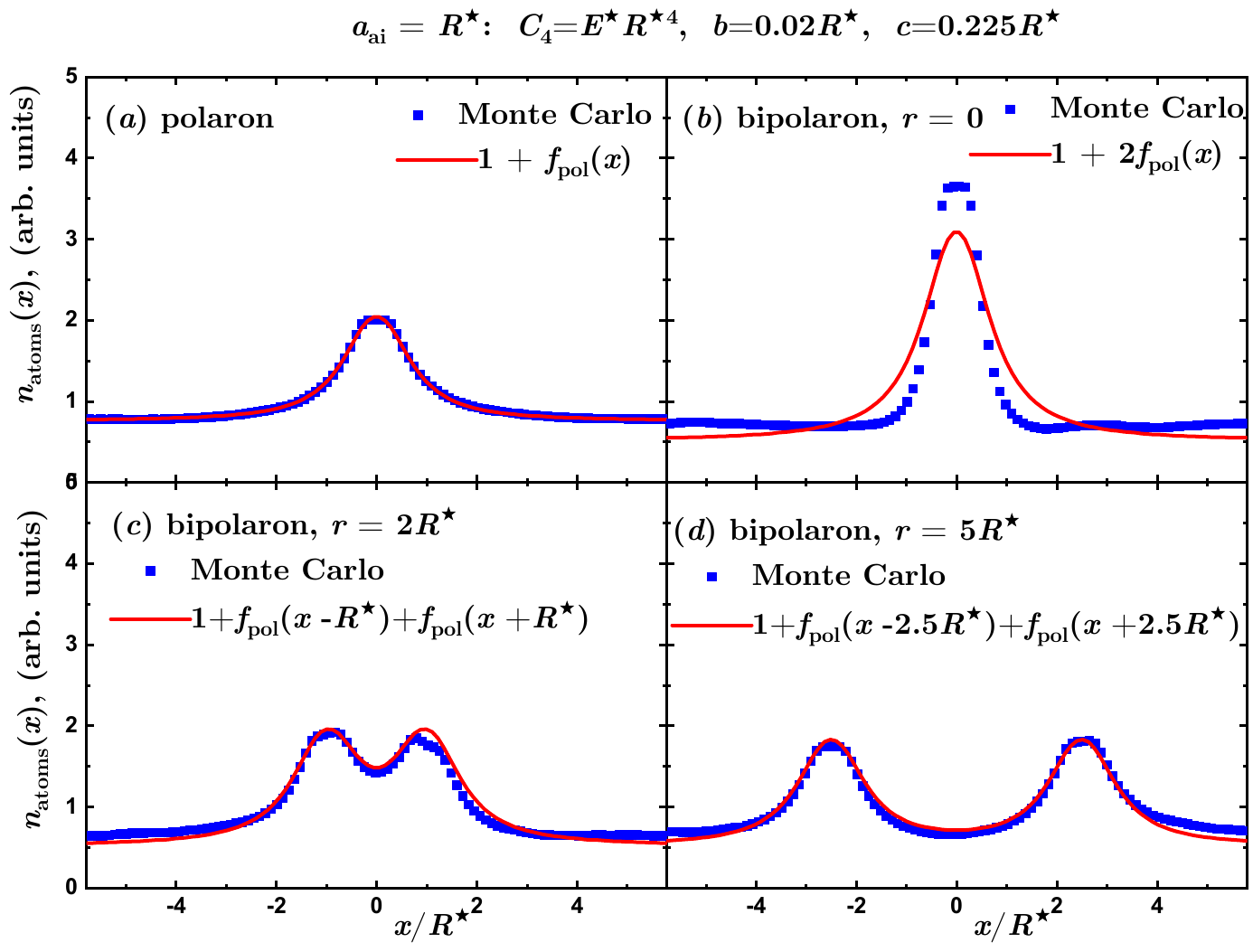}
\caption{\label{Fig:profile:a=1}
\textbf{Bipolaron density profiles in the \textrm{MBBS} regime.} Density profiles projected onto a line denoted by $x$-axis connecting two impurities in the MBBS regime (arbitrary units).
Symbols: Blue squares denote the results of QMC calculations, while the solid line is a fit to the defect in the projected atom density $n_{\mathrm{atoms}}(x)=1+f_{\mathrm{pol}}(x)$ with $f_{\mathrm{pol}}(x) = \Delta n+A/(1+Br^C)$ with $\Delta n,A,B,C$ being fitting parameters.
Panel (a): Single ion (polaron) density profile used to obtain the fit $f_{\mathrm{pol}}(x)$. Panels (b,c,d): Two-ion (bipolaron) density profile for distances between ions equal to $0,\,2R^\star,\,5R^\star$ as compared to the prediction for two non-interacting polarons located at the ion positions.
}
\end{centering}
\end{figure*}

The projected density profiles in the MBBS regime are reported in Fig.~\ref{Fig:profile:a=1}. The polaron density becomes large, differently from the weakly interacting regime presented in Fig.~\ref{Fig:profile:c=0}. Even though the Bogolyubov theory no longer describes the induced interaction potential, the bipolaron density corresponds to the one of non-interacting impurities when the distance between them is large compared to the healing length, similarly to the weakly interacting regime. Instead, for small separations, the role of the induced interaction becomes crucial, as it is manifested by significant differences in the density profile between interacting and non-interacting cases.

Summarizing the results, we notice that in all the cases we studied, regardless of the scattering length value, the induced interaction turns out to be attractive. We found that the induced interactions are consistent with a power-law Casimir decay and the magnitude of interactions strongly depends on the presence of the two-body bound state. \\


{\large \textbf{Discussion}}
\label{sec:discussion}

Our results are directly connected with the physical system consisting of a chain of trapped ions immersed in a bosonic gas. The energy spectrum of such a chain, described by phonon modes, depends on the relation between the ion separation and the Coulomb interaction. The precise knowledge of phonon frequencies is crucial e.g. for implementing quantum gates. When the size of the quantum circuit grows, even small shifts such as those coming from interaction with the background gas can lead to sizable errors. Observing such shift is also a straightforward way to study the induced interactions. Let us then estimate its typical magnitude in an experimentally relevant situation.

For mean ion separation $d \sim 1\,\mu$m ($d\simeq 13\,R^\star$ assuming Yb$^+$-Li), the Coulomb repulsion energy between two ions is $E_{\mathrm{C}} \simeq k_{\mathrm{B}}\times 16.73$ K, or $2.3\times 10^6\,E^\star$. However, the phonon mode frequencies in the trap do not depend on it directly, but rather on the second derivative of the total interaction potential calculated at equilibrium~\cite{James1998}. The lowest mode always corresponds to the center of mass motion in the trap and is unaffected by the interactions, but the second one can be tuned~\cite{James1998}. The correction is obtained simply by adding the second derivative of the induced interaction at equilibrium distance to the contribution of the Coulomb interaction. For the ion separation mentioned above and our gas parameter of $10^{-6}$, we obtain relative change of the order of $10^{-4}$ regardless of the regularized potential parameters. Note that the apparent giant enhancement of the shift close to unitarity predicted in Ref.~\cite{Ding2022} is not reproduced in our treatment, as the unitary case is not found to be dominated by the large scattering length, but rather the characteristic interaction range. Moreover, Ref.~\cite{Ding2022} uses four orders of magnitude smaller gas parameter. This enhances the gas compressibility and consequently the induced interaction, but also makes the gas more prone to large density modulations such as MBBS formation which is not included in the analytic formulas. 

Furthermore, three-body losses will strongly limit the observability of the ionic bipolarons. The expected ion lifetime due to three-body recombination in typical experimental settings is in the millisecond range unless one works with very dilute gases for which the time needed to form the (bi)polaron would be longer. However, few-body scattering calculations may in the future unravel the parameter regimes in which three-body losses are minimized.

The presence of ionic polarons can also be detected by {\it in situ} imaging of the gas. Finding a significant increase in ion-atom correlation functions as well as in the atomic density indicates the buildup of the many-body bound state. The quest for observing the induced interactions is more subtle, as typically the direct Coulomb repulsion would be too strong to allow for the formation of bipolaronic bound states or scattering resonances. However, in this case, the study of atomic density looks promising. One could measure e.g. the deviation of the density profile from the double Gaussian peak describing separated noninteracting ionic polarons. 

Finally, we note that our results are also relevant for other systems with long-ranged interactions competing with the length scales of the medium where the direct interaction potential may not be dominating, such as Rydberg-dressed mixtures.\\ 

In conclusion, we have demonstrated that impurities strongly interacting with the host medium not only experience an effective interaction that can lead to the formation of bipolaronic states, but also dramatically modify the gas properties. Because of the strong modification of the gas density profile around the ion(s), perturbative methods based on dressing the cloud with Bogolyubov excitations and neglecting the bound state occupation do not fully capture the description of the induced potential which features a kink close to the local density maximum. This indicates that ab-initio many-body simulations are of paramount importance for studying long-ranged impurity-bath interactions. 

While the magnitude of the effective potential is vast compared to the gas energy scales, it is still much smaller than the Coulomb repulsion between the ions. Notwithstanding, it leads to shifts in the phonon mode energies of an ionic chain compared to the vacuum case that can be experimentally observed. This fact might be relevant for quantum technologies based on trapped ions, where the phonon modes are exploited as a ``quantum bus" to mediate interactions between spatially separated quantum bits. As we discussed in the introduction, a cold atomic ensemble could be exploited to keep the ions cold to aim at fault-tolerance quantum computation for long times. Our investigations show that the phonon modes can be affected by the presence of the gas. At the same time, however, our findings prove that the phononic shift can be controlled by tunning both the number of two-body atom-ion bound states and the ion-atom scattering length, providing thus,  an additional tool for tuning the phonon modes of an ion crystal. \\

In the future, it will be interesting to investigate the impact of the ion motion degrees of freedom both on the ground state and transport properties, especially in the many-body bound state regime. Furthermore, the role of larger ionic chains and the possibility of multipolarons states~\cite{Yegovtsev,Ardila2021-2} in hybrid atom-ion systems may also be an interesting path to explore. Finally, an important issue is finite temperature effects and how thermal fluctuations affect our findings. In this regard, we note that Monte Carlo techniques can be used as well, as has been shown in a recent study on the neutral Bose polaron~\cite{JordiPRL2021}.\\

{\large \textbf{Methods}}\\
{\textbf{Numerical Method} --} We employ the diffusion Monte Carlo method which computes the ground state energy of Hamiltonian~(\ref{eq:H}) by propagating the many-body Schr\"odinger equation in imaginary time. The boson-boson interaction is modelled by soft-spheres with a diameter $R_{\mathrm{SS}}$ small as compared to the atom-ion range, i.e., $R_{\mathrm{SS}}=0.1R^\star$, whereas the height is adjusted to have a small value of the three-dimensional $s$-wave scattering length $a_{\mathrm{bb}}=0.02R^\star$. The guiding wave function is written in a pair product form $\Psi = \prod_{i<j} f_{\mathrm{BB}}(r_{ij}) \prod_{i,\alpha} f_{\mathrm{BI}}(r_{i\alpha})$, similarly to the one used in Ref.~\cite{Gregory2021}. It consists of boson-boson and boson-ion Jastrow pair-product terms, each one constructed in such a way that the two-body scattering at small distances matches the phononic long-range asymptotic~\cite{ReattoChester67}. Specifically, calculations are performed for $N=200$ bosons in a box with periodic boundary conditions and with $N_{\mathrm{I}} = 2, 1, 0$ ions. We consider dilute densities with the gas parameter equal to $na_{\mathrm{bb}}^3 = 10^{-6}$. In that case, the atomic chemical potential is small as compared to the typical ion energy, $\mu_{\mathrm{bb}} = 4\pi\hbar^2 n a_{\mathrm{bb}}/m = 0.0314 E^\star$, and the healing length is larger than the characteristic interaction length, $\xi = 4R^\star$.\\

The energy shift due to the interaction between two ionic polarons mediated by the bath is computed as
\begin{equation}
E^{pol-pol} = E(N;2)-2E(N;1)+E(N;0),
\end{equation}
where $E(N;N_{{\mathrm{I}}})$ denotes the ground-state energy of the system containing $N$ atoms and $N_{\mathrm{I}}$ ions. 
In the case of neutral impurities, this value is on the order of the single polaron energy $E_{\mathrm{n}}=\frac{\hbar^{2}}{2m}\left(6\pi^{2}n\right)^{2/3}\sim E^{\star}$ for very large values of 
of the neutral impurity-boson scattering length $a_{\mathrm{ab}}$. The induced interaction is attractive regardless the sign of $a_{\mathrm{ab}}$~\cite{georg2021} as well as for the case where a two-body bound-state for $a_{\mathrm{ab}}<0$ 
does not exist. For the atom-ion compound system we consider both two-body bound and scattering states.\\

\subsection*{Data availability}

The data generated in this study have been deposited in the figshare database under  the DOI: https://doi.org/10.6084/m9.figshare.22134527.v1

\subsection*{Code availability} 

The code that supports the plots within this paper is available from the first author upon reasonable request.

\subsection*{Acknowledgements}
We acknowledge discussions with Georg Bruun on his related work~\cite{Ding2022}. 
This work is supported by the project NE 1711/3-1 of the Deutsche Forschungsgemeinschaft, the Polish National Agency for Academic Exchange (NAWA) via the Polish Returns 2019 programme, by the Secretaria d'Universitats i Recerca del Departament d'Empresa i Coneixement de la Generalitat de Catalunya, co-funded by the European Union Regional Development Fund within the ERDF Operational Program of Catalunya (project QuantumCat, ref.~001-P-001644), Grant PID2020-113565GB-C21 funded by MCIN/AEI/10.13039/501100011033, the Spanish MINECO (FIS2017-84114-C2-1-P), and the Secretaria d'Universitats i Recerca del Departament d'Empresa i Coneixement de la Generalitat de Catalunya within the ERDF Operational Program of Catalunya (project QuantumCat, Ref. 001-P-001644). LA  acknowledges support of the Deutsche Forschungsgemeinschaft (DFG, German Research Foundation) under Germany’s Excellence Strategy– EXC-2123 QuantumFrontiers– 390837967, and FOR 2247.

\subsection*{Author contributions}

G.E.A. and L.A.P.A. performed the Monte Carlo simulations with input from the other authors. G.E.A, L.A.P, K.J and A.N proposed the research project and contributed equally to the analysis of the results and to the writing of the manuscript.

\subsection*{Competing interests}
The authors declare no competing interests.


\end{document}